\documentclass[aps,pre,twocolumn,showpacs,groupedaddress]{revtex4}  
\usepackage{graphicx}  
\usepackage{dcolumn}   
\usepackage{bm}        
\usepackage{amssymb}   

\begin{document}

\title{High-order synchronization, transitions, and competition among\\
Arnold tongues in a rotator under harmonic forcing}

\author{David Garc\'\i a-\'Alvarez}
\email{d.garcia-alvarez@lancaster.ac.uk}

\author{Aneta Stefanovska}
\author{Peter V.E. McClintock}

\affiliation{
\centerline{Department of Physics, Lancaster University, Lancaster LA1 4YB, United Kingdom}
}

\date{March 12, 2008}


\begin{abstract}
We consider a rotator whose equation of motion for the angle
$\theta$ consists of the zeroth and first Fourier modes.
Numerical analysis based on the trailing of saddle-node
bifurcations is used to locate the $n$:$1$ Arnold tongues where
synchronization occurs. Several of them are wide enough for
high-order synchronization to be seen in passive observations.
By sweeping the system parameters within a certain range, we
find that the stronger the dependence of $\dot\theta$ on
$\theta$, the wider the regions of synchronization. Use of a
synchronization index reveals a vast number of very narrow
$n$:$m$ Arnold tongues. A competition phenomenon among the
tongues is observed, in that they ``push'' and ``squeeze'' one
another: as some tongues widen, others narrow. Two mechanisms
for transitions between different $n$:$m$ synchronization
states are considered: slow variation of the driving frequency,
and the influence of low-frequency noise on the rotator.
\end{abstract}

\maketitle

\section{Introduction}

When two or more oscillatory processes are coupled, there exists
the possibility of their becoming synchronized. Where their
autonomous frequencies are different but close, synchronization is
understood as the adjustment of those frequencies as a result of
coupling. Even when such systems operate on different timescales,
synchronization may still appear as an adjustment of their
frequencies to an integer ratio, an effect known as {\it
high-order synchronization} or synchronization of order $n$:$m$.

Synchronization of order $n$:$m$ has been extensively studied,
both experimentally and theoretically (cf. \cite{arkadybook} for
a review). For instance, Simonet and co-workers \cite{simonet}
investigated a ruby nuclear magnetic resonance laser with
delayed feedback. The undriven laser exhibited periodic
oscillations of light intensity at a frequency $\nu_0$ $\simeq$
40 Hz. An external periodic voltage, either sinusoidal or
square-wave, was then added in the feedback loop. In both
cases, synchronizations of different order $n$:$m$ were
observed. Another example is the electrical rotator
consisting of a Josephson junction, shunted by a capacitor, and
fed with a constant external current. Experiments show that
this system can be synchronized when a periodic driving current
is applied, or where two junctions are coupled
\cite{jainjosephson}. There are many additional examples.

Synchronization has also been observed extensively in biology.
One example is the cardiorespiratory system, considered in the pioneering works by  Kenner {et al}, Hildebrandt and Raschke \cite{raschke}.  Sch\"afer and
co-workers \cite{schafer} proposed the synchrogram as a tool to visualize cardiorespiratory synchronization.
 When plotting the instantaneous
respiratory phase at the occurrence of a heartbeat versus time,
 they found horizontally striped plots
for some subjects, thereby revealing $n$:$1$ synchronization
between heart and respiration. Toledo et al \cite{toledo}
showed that the probability of such synchronization happening
by chance was extremely small. In measurements on anaesthetised
rats, Stefanovska et al \cite{anetarats} observed lengthy
synchronization epochs, and transitions from one ratio to
another. They suggested that such transitions might be useful
in monitoring depth of anaesthesia.

There are several nonlinear models yielding $n$:$m$
synchronization. Arnold proposed \cite{arnold} a map of the
circle into itself, obtaining synchronization ``tongues'' and
calculating their widths in the approximation of small
coupling. In fact, any orientation-preserving homeomorphism
$h$:$S^1\rightarrow S^1$ of the circle into itself presents
such regions of $n$:$m$ locking \cite{katok, boyland}.
Integrate-and-fire models  provide several examples of Arnold
tongues \cite{tiesinga}. High-order synchronization regions
were also obtained by Glass and Sun \cite{glasssun} for an
impulse-driven Poincar\'e oscillator. Schilder and Peckham \cite{tonguesvanderpol} treated Arnold
tongues numerically, and they obtained tongues for the system of two
coupled Van der Pol oscillators; here the tongues are quite
narrow, so that the probability of locking in a real, noisy
system, would therefore be very small. Simonet and co-workers
proposed a model \cite{simonet} that reproduced the
synchronizations observed in the laser. For cardiorespiratory
synchronization, Kotani et al \cite{kotani} developed a model,
based on those of DeBoer {et al} \cite{deboer} and Seidel and
Herzel \cite{seidelherzel}, which is supported on both
physiological and mathematical principles. The model involves a
somewhat complicated system consisting of several oscillators
and interactions, and includes some non-analytic parts
(integrate-and-fire). Because of technical difficulties
encountered when tackling nonlinear models, the deep mechanisms
through which synchronization takes place are not yet
understood in general, so that there is still no way of
predicting which equations, and which values of their
parameters, will or will not yield synchronization.

The present paper has two main purposes. First we present a
systematic study of high-order synchronization in a
particularly simple system, a rotator under harmonic forcing,
for which we can establish the roles played by each of its
parameters in synchronization. Although we study a specific
system, and although the deep mechanisms responsible for
synchronization are not unveiled in this paper, we report below
two results that we believe will be useful in the quest for
those deep mechanisms: we show that the main Arnold tongues are
wider when $\dot\theta$ depends more strongly on $\theta$; and
that competition occurs between synchronization regions. Our
second purpose is to discuss and explore two possible
mechanisms giving rise to an extensively observed phenomenon:
transitions between the different $n$:$m$ synchronization
ratios. The mechanisms considered here arise from time
variability. We will also discuss briefly why time variability
hinders the analysis of synchronization in experimental data.

The paper is organized as follows: In Sec.\ \ref{sec:system} we
introduce the simple rotator whose synchronization properties
are to be considered. Sec.\ \ref{sec:synctongues} discusses how
its regions of $n$:1 and $n$:$m$ synchronization are identified
and reports the main results obtained under stationary
conditions, including the observation of competition between
the tongues. Time variability and its effect on transitions
between different synchronization states is discussed in Sec.\
\ref{sec:transitions}. Sec.\ \ref{sec:conclusions} summarizes
the main conclusions.

\section{The system}\label{sec:system}

The generic equation for the angle $\theta$ of a rotator
without external interaction is
$$\dot\theta=f(\theta),$$
where $f$ is a $2\pi$-periodic function \cite{foot1}.
Therefore, such a system can be studied systematically by
considering functions $f$ up to a certain number $k$ of
harmonics, and allowing a bigger number $k+1$ of harmonics at
the next stage of the study. In this paper, we start the study
for a function consisting of the zero harmonic (the ``constant
force'', thanks to which a rotator has the features of a
self-sustained oscillator) and the first harmonic. By means of
a translation in the value of $\theta$, it can always be
written as
\begin{equation}\label{ourrotator}
 \dot\theta=a_0+a_1\, \cos\theta,
\end{equation}
with $a_0$ and $a_1$ constants, where (\ref{ourrotator}) is an
Adler-type equation \cite{adler}. We assume $a_0>|a_1|$ so
that, in the absence of interaction (nothing is added to
(\ref{ourrotator})), the angle $\theta$ continuously increases.
For $a_0$ much bigger than $a_1$, $\theta$ increases at an
almost constant rate. For $a_1$ close to $a_0$, however,
$\dot\theta$ varies strongly with $\theta$. We show that
(\ref{ourrotator}) can synchronize to an external forcing,
exhibiting a wide variety of Arnold tongues, and we discuss the
processes that may be responsible for transitions between
different $n$:$m$ synchronization ratios.

The equations for the overdamped pendulum, and the overdamped
Josephson junction, are of just this type. The equation of
motion of a pendulum driven by a constant torque $K$ is
described by
\begin{equation}\label{pendulum}
 \ddot\Psi+\gamma\, \dot\Psi+\kappa^2\, \sin\Psi=\frac{K}{I},
\end{equation}
where $\kappa$ is the frequency of small oscillations,
$\gamma>0$ is the damping constant, and $I$ is the moment of
inertia.

In a resistively shunted Josephson junction, the current is a
sum of three contributions: a superconducting current $I_c\,
\sin\Psi$, a current $V/R$ through the resistance, and a
capacitance current $\dot V\, C$. The parameter $I_c$ is called
the critical supercurrent of the junction. The relationship
between the potential $V$ and the ``angle'' $\Psi$ is given by
the Josephson formula
\begin{equation}
 \dot\Psi=\frac{2e}{\hbar}V,
\end{equation}
where $e$ is the electronic charge. Summation of the three
terms yields the equation for a junction fed with an external
current $I$
\begin{equation}\label{josephsonjunction}
I= I_c\, \sin\Psi+\frac{\hbar}{2e\, R}\frac{d\Psi}{dt}+\frac{C\hbar}{2e}\frac{d^2\Psi}{dt^2}.
\end{equation}
Equations (\ref{pendulum}) and (\ref{josephsonjunction})
coincide. In the overdamped limit, when the term with the
second derivative can be neglected -- in the case of the
Josephson junction, this means that there is no capacitor in
the circuit -- these equations reduce to the one that we study
in this paper (\ref{ourrotator}). Thus the results that we
obtain below will be applicable to the overdamped pendulum and
the overdamped Josephson junction.

As (\ref{ourrotator}) is analytically integrable, we find that
the frequency for the non-interacting rotator is
\begin{equation}\label{isofrequency}
 \nu_0=\frac{1}{2\pi} \sqrt{a_0^2-a_1^2}.
\end{equation}

We now consider the effect of an external harmonic force on the
rotator
\begin{equation}\label{rotatorforced}
 \dot\theta= a_0 + a_1 \cos\theta+B\sin(\omega\, t),
\end{equation}
with $B\geq 0$. This equation applies to a number of situations
in nature. In the case of a Josephson junction, Eq.\
(\ref{rotatorforced}) describes when the system is fed with a
continuous intensity plus a harmonic one. In relation to
cardiorespiratory synchronization, (\ref{rotatorforced}) can be
regarded as a very simple model in which the rotator
(\ref{ourrotator}) models the heart, and the addition of the
harmonic component models its interaction with respiration
\cite{foot2}. Eq.\ (\ref{rotatorforced}) has been studied
\cite{arkadybook} for large amplitudes $B$ of the harmonic
component. However, we are interested here in the Arnold
tongues down to very small driving amplitude, in order that we
can also apply the results to the weakly coupled rotator.

Eq.\ (\ref{rotatorforced}) defines a circle map with the
following prescription: let us call  $t_i$  the time at which
the external driving is at its $i^\mathrm{th}$ maximum. We
define $\theta_i\equiv\theta(t_i)$. As (\ref{rotatorforced}) is
a first-order differential equation and we have an initial
condition $\theta(t_i)$, we could integrate to obtain
$\theta(t)$. Let $t_{i+1}$ be the time at which the external
driving is at its $i+1$ maximum, and let
$\theta_{i+1}\equiv\theta(t_{i+1})$. That is how we have the
map $h$:$S^1\rightarrow S^1$ defined as
$\theta_{i+1}=h(\theta_i)$. We would therefore expect the
existence of Arnold tongues \cite{katok, boyland}.

\section{Synchronization and Arnold tongues}\label{sec:synctongues}

\subsection{Synchronization of a rotator driven by an external periodic force}

Now we introduce the concept of $n$:$m$ synchronization that
will be used in this work. Suppose we have a rotator  driven by
a $T_E$-periodic external action
\begin{eqnarray}
&&\dot\theta = f(\theta)+g(t),\qquad\mathrm{with} \nonumber\\
&&f(\theta+2\pi)=f(\theta)\ \forall\theta,\qquad g(t+T_E)=g(t)\ \forall t.\nonumber
\end{eqnarray}
Then, any $T$-periodic motion of $\theta$,
$\theta(t+T)=\theta(t)\ \mathrm{mod}\ 2\pi$
, must have a period that is a multiple of the driving period.
Let $m\in\mathbb{N}$ be such that $T=m\, T_E$. Let
$n\in\mathbb{N}\cup\{0\}$ be the number of times that the angle
crosses $\theta=0$ mod $2\pi$ with $\dot\theta>0$ in one of its
periods $T=m\, T_E$. We then say that the rotator is $n$:$m$
synchronized to the external driving $g$. The synchrogram
consists then  of \emph{horizonal lines}, albeit not
necessarily equally spaced. Note that any periodic motion of
the forced rotator automatically implies $n$:$m$ locking to the
external action for some $n$ and $m$.

\subsection{Saddle-node bifurcation}

Putting the above definition of $n$:$m$ synchronization into
mathematical terms, we say that the rotator gets synchronised
$n$:$m$ to the external force if there exists a stable root
(zero) for the function
\begin{equation}\label{poincarereturn}
 h_{nm} (\theta) \equiv h^m (\theta) - 2  \pi n -\theta,
\end{equation}
where $h^m$ stands for the return map $h$ composed with itself
$m$ times. In general, if $h_{nm}$ has two or more roots, there
is at least a stable fixed point. Regardless of the initial
condition, the trajectory is attracted towards a stable fixed
point. The function $h_{nm}$ depends, of course, on the
parameters of (\ref{rotatorforced}), so it changes as we vary
the external driving frequency $\omega$. At the moment of
transition from two zeros to no zero, there is only one root,
on which $h_{nm}$ is tangent to the horizontal axis,  this
single zero is   a half-stable fixed point. At this moment, a
\emph{saddle-node bifurcation} takes place.

Obtaining the borders of the Arnold tongues therefore involves
retrieving the two driving frequencies at which the saddle-node
bifurcation takes place for different values of the driving
amplitude $B$. For this purpose a \emph{continuation software}
was written in C. The first step was to obtain the Poincar\'e
return map $h$, so the interval $[0, 2\pi]$ was divided into
many points. At each of these, the function $h^m$ was obtained
by integration of the differential equation
(\ref{rotatorforced}) with the fourth-order Runge-Kutta method,
thereby obtaining $h_{nm} (\theta) = h^m (\theta) - 2 \pi
n-\theta$. For given values of $B$ and $\omega$, we know that
we are inside the tongue if $h_{nm}$ has two or more roots, and
outside if $h_{nm}$ has no zeros. Hence, for a given value of
$B$, we can trace the left and right values of $\omega$ at
which the saddle-node bifurcation takes place, up to the
desired precision. We start with zero driving amplitude, $B=0$;
for this value the tongue consists of only one point $\omega=m\
\omega_0/n$ -- this will not be forced when we study the
bifurcations in the ``flexible'' way for very narrow tongues,
see Section \ref{sec:competition}. For the next value of the
coupling (positive but close to 0), we start from $\omega=m\
\omega_0/n$ and look for the two bifurcation points. Then, for
sequentially increasing values of $B$, we first guess
approximately the bifurcation points on the left and right
boundaries by linear extrapolation from the two former
bifurcation frequencies in each case. Starting from this
guessed value, the programme looks for the correct bifurcation
point. In some tongues and for small values of $B$, the
synchronization region is so narrow that the programme cannot
retrieve the bifurcation points. In such cases, the programme
skips this value of $B$, goes to the next $B$, and sets the
starting point (the ``guessed value'' of the bifurcation
frequency) based on the former bifurcation points that were
successfully retrieved.

\begin{center}
 \begin{figure}[t!]
\vspace{-2em}
$$  \begin{array}{c}
\hspace{-2em}
 \includegraphics[height=0.53\textwidth, width=0.3\textheight, angle=270]{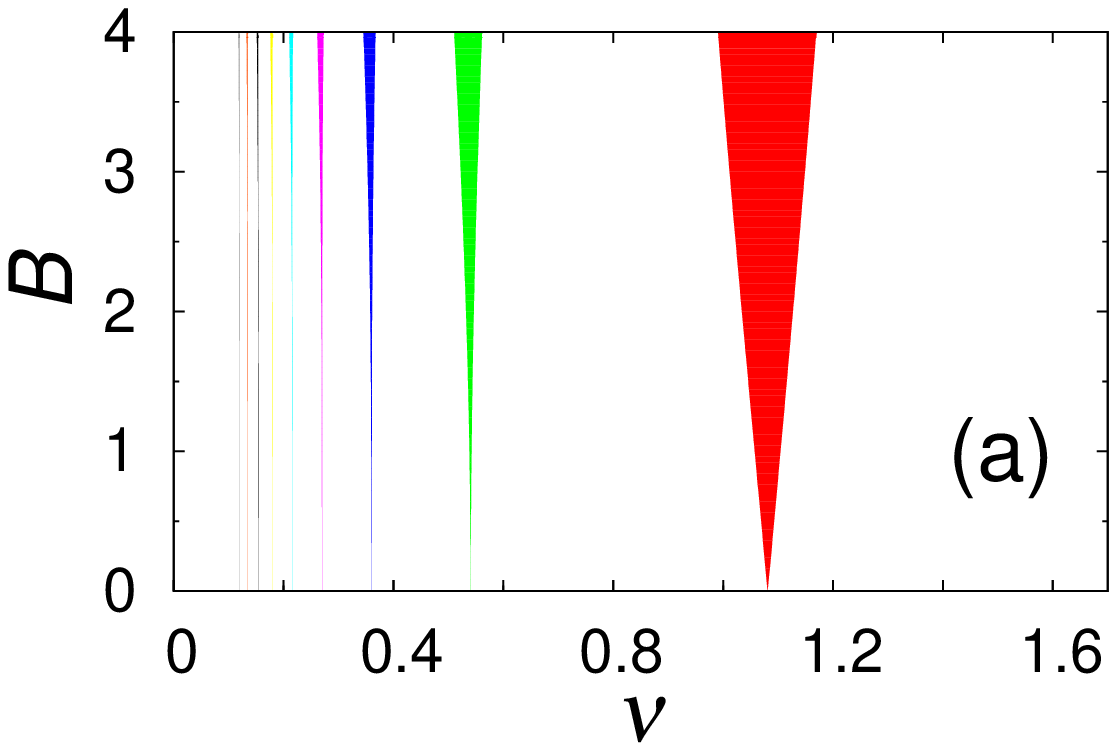}\\ \vspace{-5em}\\
\hspace{-2em}
 \includegraphics[height=0.53\textwidth, width=0.3\textheight, angle=270]{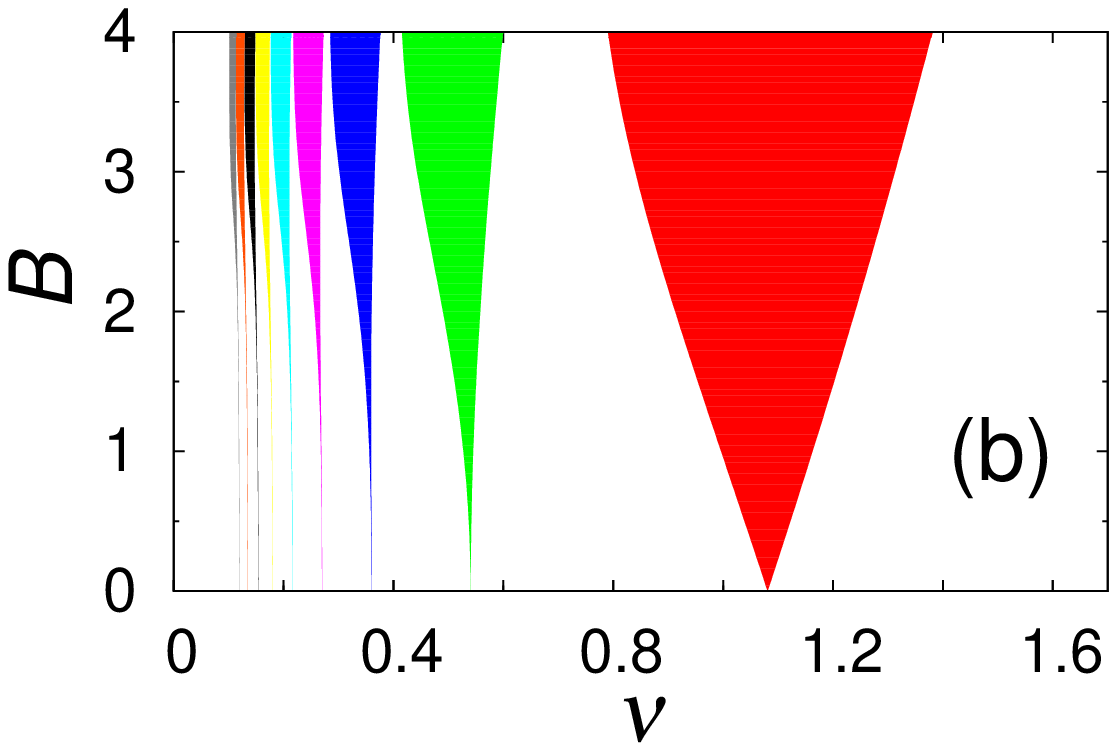}\\  \vspace{-5em}\\
\hspace{-2em}
 \includegraphics[height=0.53\textwidth, width=0.3\textheight, angle=270]{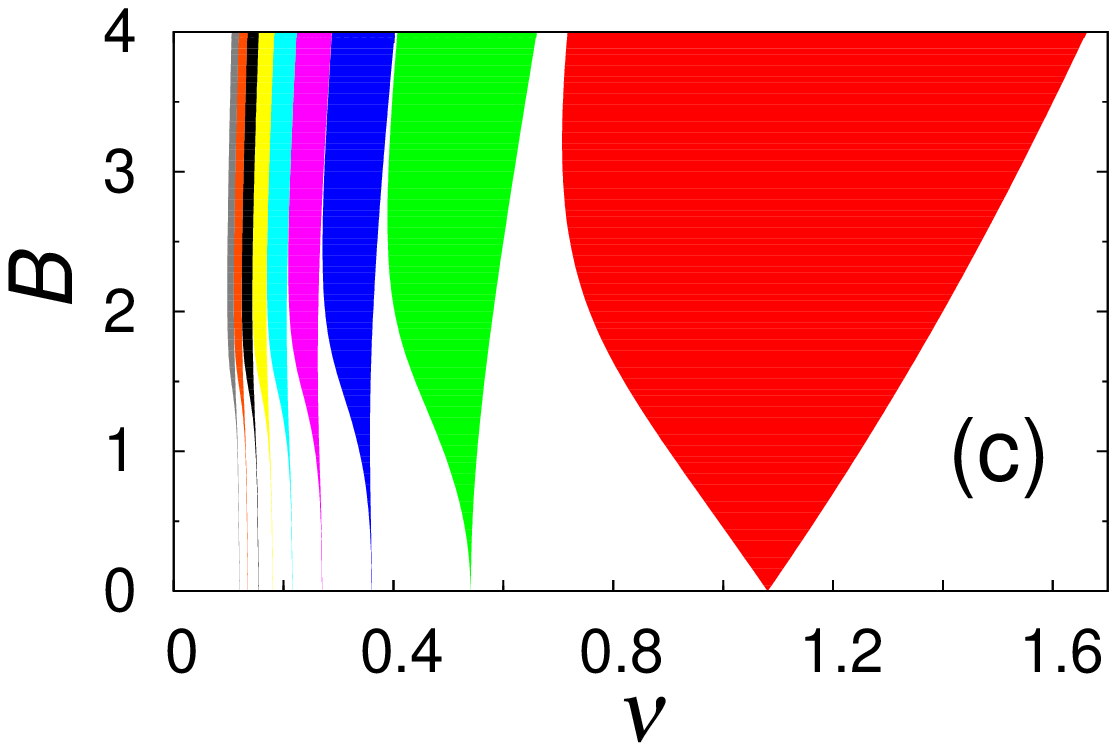}\\ \vspace{-3em}
  \end{array} $$
\caption{\label{fig:3tongues} (colour online) Arnold tongues for
the externally-driven rotator (\ref{rotatorforced}), with
parameters $(a_0, a_1)$ equal to: (7.08116, 2) in (a); (9.85714,
7.14286) in (b); and (16.46637, 15) in (c). The frequency $\nu$ and amplitude $B$ of the
external driving are plotted on the abscissa and ordinate axes respectively. The
tongues are $n$:1, with $n$ increasing from 1 to 9 as we move from
right to left.}
 \end{figure}
\end{center}

\vspace{-3em}

\subsection{The \textit{n}:1 synchronization regions}

\begin{center}
 \begin{figure*}[t!]
$$  \begin{array}{cc}
 \includegraphics[height=0.5\textwidth, angle=270]{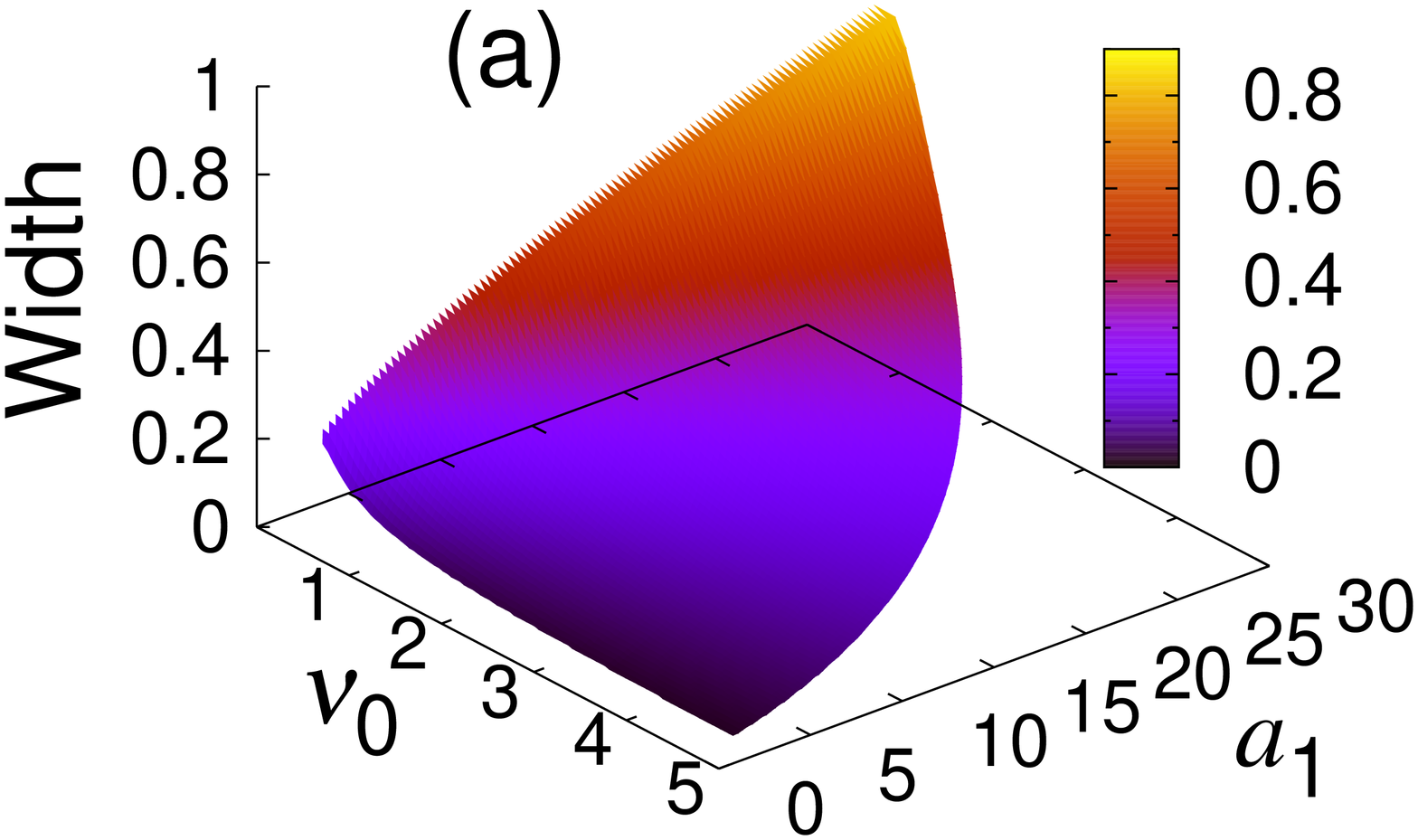} &
\includegraphics[height=0.5\textwidth, angle=270]{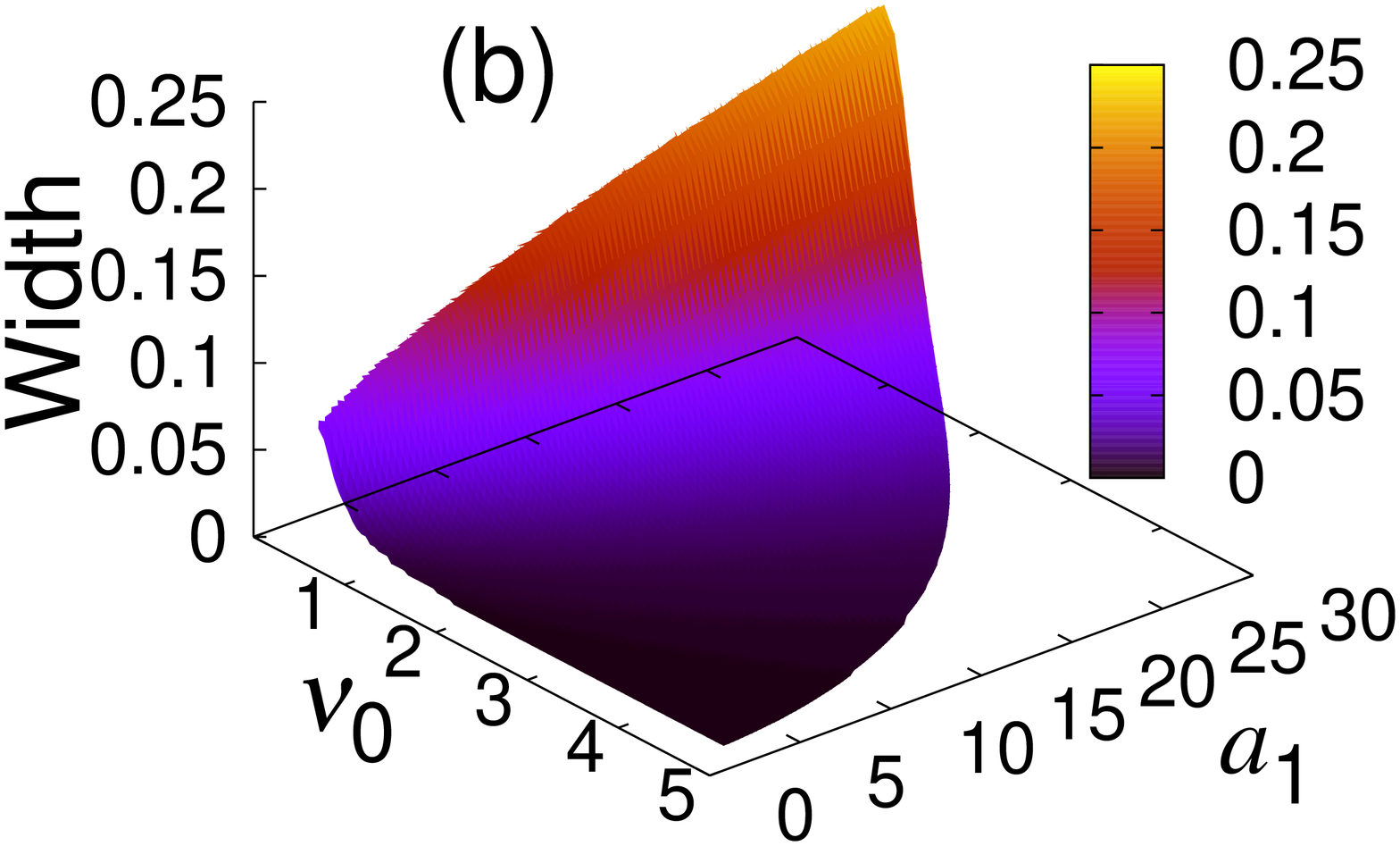} \\
 \includegraphics[height=0.5\textwidth, angle=270]{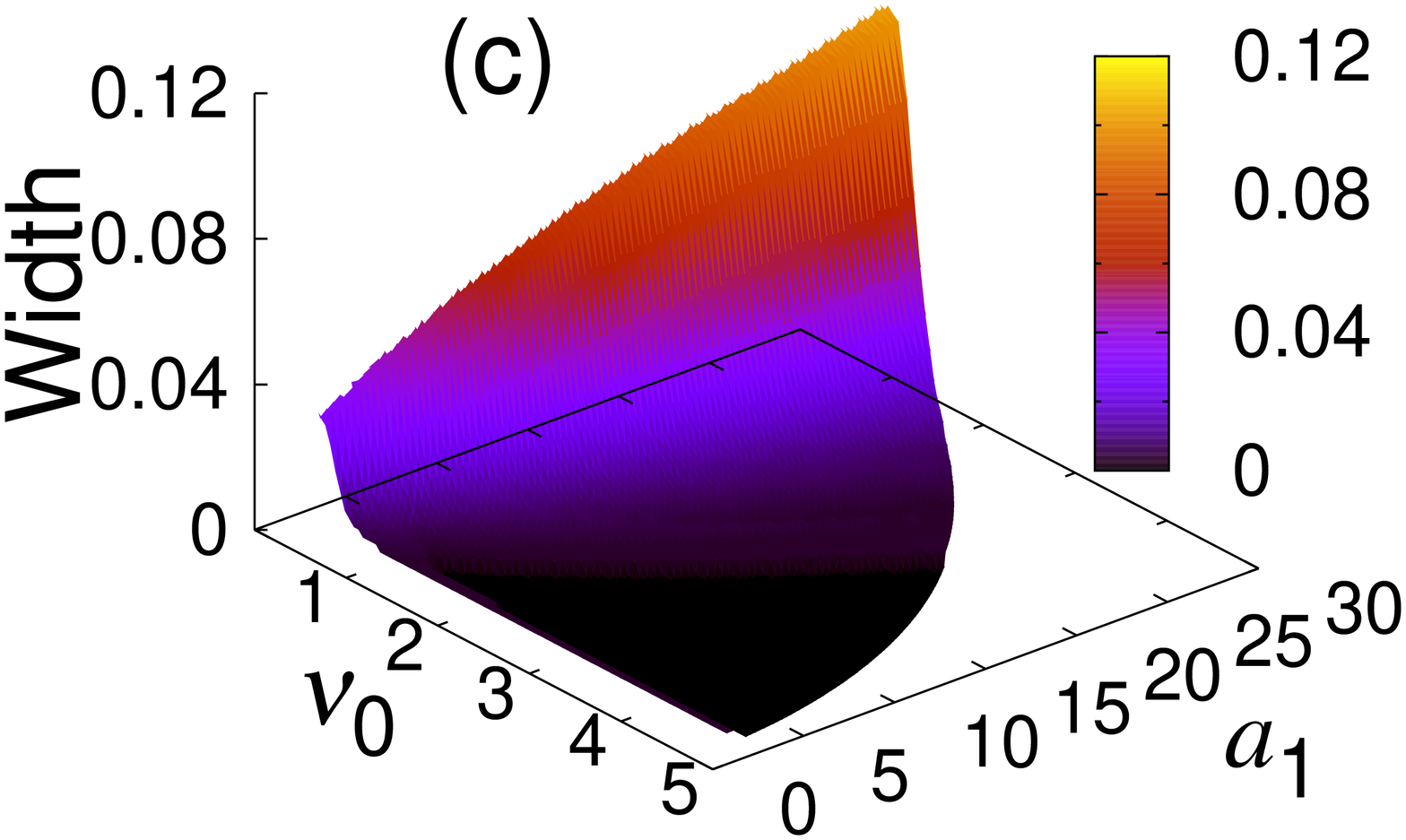} &
\includegraphics[height=0.5\textwidth, angle=270]{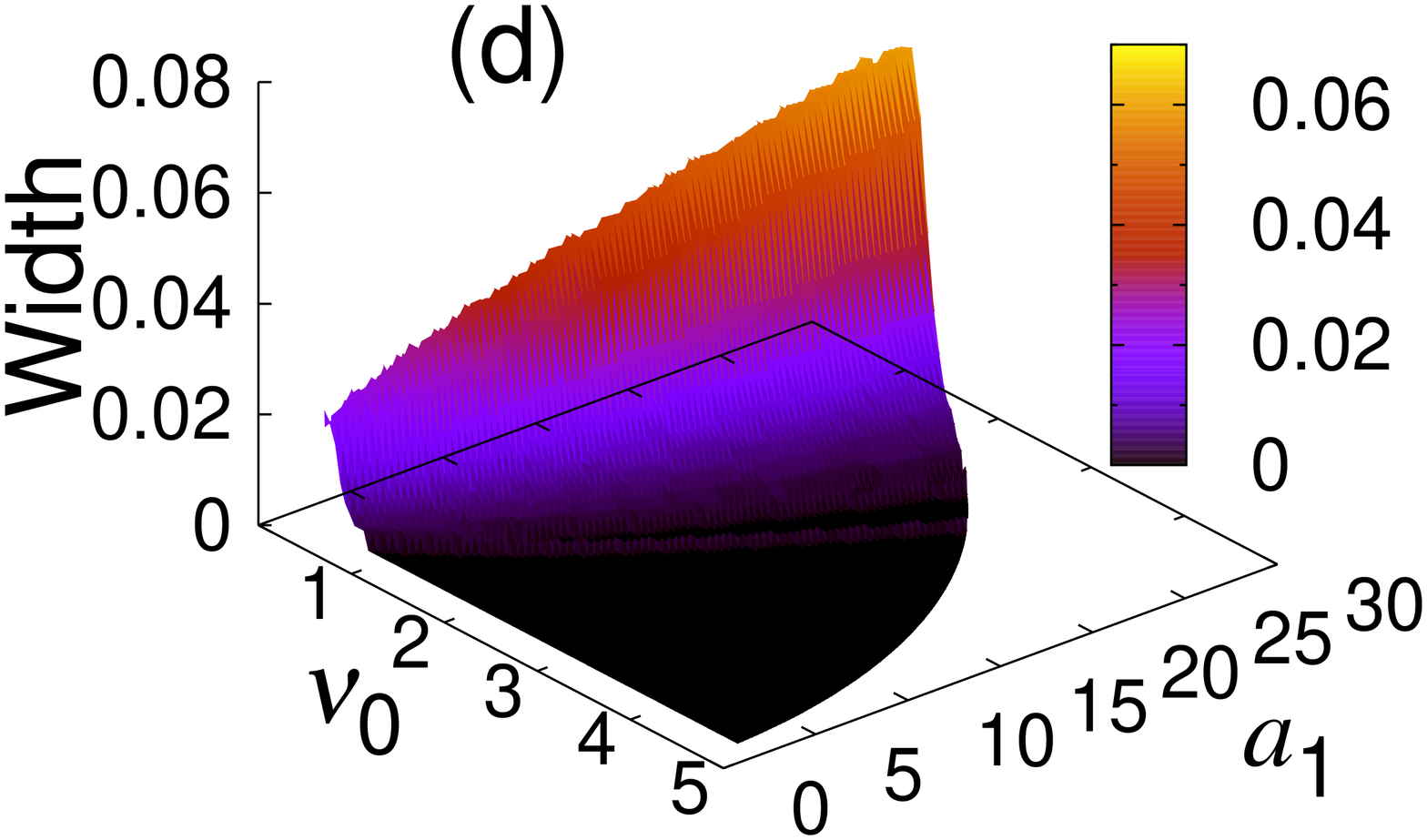}
  \end{array} $$
\caption{\label{fig:widths} (color online) Widths of the Arnold
tongues as functions of parameters $a_1$ and $\nu_0$: (a) 1:1;
(b) 1:2; (c) 1:3; and (d) 1:4. They are expressed in units of
the driving frequency $\nu$, at driving amplitude $B=1.5$. Note
the differences in ordinate scale.}
 \end{figure*}
\end{center}

We obtained the first $n$:1 tongues for three combinations of
parameters $(a_0, a_1)$ that, according to (\ref{isofrequency}),
yield the same autonomous frequency for the rotator $\nu_0\simeq
1.0811$: (7.08116, 2), (9.85714, 7.14286) and (16.46637, 15). Note
that dimensionless units are used for all variables throughout
this paper. The results are plotted in Fig.\ \ref{fig:3tongues}.
The tongues with $m>1$ are very narrow and the programme fails to
retrieve them by the method of seeking the saddle-node
bifurcations. Later we will use other less accurate methods for
the high-order tongues. As often reported in the literature
\cite{arnold}, the widest tongues are those for $m=1$. For our
system (\ref{rotatorforced}), as $n$ increases, the widths of the
$m=1$ tongues decrease and the tongues become closer to one
another. A point of interest here is that there are several Arnold
tongues that are wide enough to be seen, not only by seeking them
in a tuned experiment, but also through passive observations in
the presence of noise, e.g.\ the cardiovascular system
\cite{schafer, anetabracic, anetarats}.

In our second parameter set, $a_1$ is bigger than in the first
set, and longer again in the third set compared to the second one.
Thus, the value of $\dot\theta$ depends much more strongly on
$\theta$ as we move from (a) to (c) in Fig.\ \ref{fig:3tongues},
although the frequency $\nu_0$ remains the same. The effect on the
Arnold tongues is immediately evident on comparing the three plots
in Fig.\ \ref{fig:3tongues}: the tongues become markedly wider,
thereby favouring synchronization. We may therefore conjecture
that a strong dependence of the instantaneous frequency on angle
(for the uncoupled rotator) may help high order synchronization.

In order to check this conjecture more thoroughly, we fix the
value of the driving amplitude $B$ to 1.5, and we compute the
widths of the tongues at $B=1.5$ for different values of the
parameters $a_0$ and $a_1$. We swept $a_0$ from 2 to 30, and
$a_1$ from 2 to (almost) $a_0$. In order to compute the width
of the tongues, the code was modified, so that the two driving
frequencies at which the saddle-node bifurcations take place
were computed only for this value of $B$. Starting from
$\omega=m\, \omega_0 /n$, we trailed the bifurcation points
first with small precision, then with a greater one,
recursively up to the desired precision. We plot in Fig.\
\ref{fig:widths} the widths of the tongues at $B=1.5$ versus
the autonomous frequency $\nu_0$ and the parameter $a_1$ (to
which $a_0$ is related through (\ref{isofrequency})). In the
figure we see that, for the same autonomous frequency $\nu_0$,
the widths of the tongues dramatically increase with $a_1$,
confirming the conjecture that a strong dependence of
$\dot\theta$ on $\theta$ (for the uncoupled rotator) favours
synchronization. Regarding this conjecture, it would be very
interesting to add higher harmonics to (\ref{ourrotator}) and
to study how the widths of the tongues for the driven rotator
change. Nevertheless, in this case, the frequency $\nu_0$ of
the isolated rotator is not an elementary function, so it would
perhaps be more convenient to calculate this frequency
numerically.

\subsection{Use of the synchronization index}

As mentioned above, the tongues with $m>1$ are very narrow and
the programme fails to retrieve them by the method of seeking
the saddle-node bifurcations. In order to obtain the
synchronization regions in such cases, we have been obliged to
use other (albeit less accurate) approaches. The first one is
via a synchronization index: we generated data by numerical
integration of Eq.\ (\ref{rotatorforced}) with the fourth-order
Runge-Kutta method; and we analysed them as though they were
experimental data, by computing their synchronization index.
With the definition of synchronization we mentioned at the
beginning, it is natural to quantify it by how close the
following index \cite{kotani} is to zero 
\begin{equation}\label{synchindex}
 H_{mn}=\frac{1}{2\pi\, m\, (L-n)} \left| \sum_{i=n+1}^L (\Phi_i-\Phi_{i-n}-2\pi m)\right|,
\end{equation}
where $L$ is the total number of crossings of $\theta=0$ mod
$2\pi$ with $\dot\theta>0$, and $\Phi$ is the phase of the
external action when the $i^\mathrm{th}$ crossing occurs.

\subsection{The \textit{n}:\textit{m} synchronization regions}

To illustrate the $n$:$m$ synchronization regions, we again
plot the driving frequency $\nu$ on the horizontal axis, and
the coupling or driving amplitude $B$ on the vertical axis. The
horizontal axis was divided into 5,000 points, and the vertical
one into 30 points. For each of the 150,000 corresponding pairs
of points, the Eq.\ (\ref{rotatorforced}) was integrated. The first 60 zero-positive
crossings were discarded to remove transients. The maximum
accepted value of the index $H_{mn}$ (\ref{synchindex}) for a
point to be considered as belonging to the $n$:$m$ Arnold
tongue was 0.0005.

\begin{center}

\begin{figure}
\includegraphics[width=0.48\textwidth]{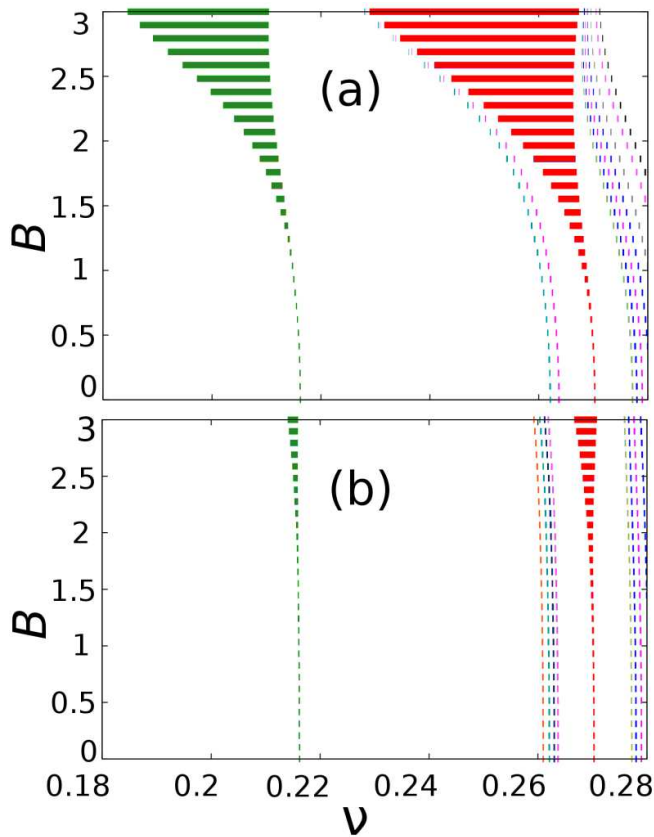}
\caption{\label{fig:tonguesdetailed} (colour online) Detail of the
synchronization regions inside a small interval.  The parameters
$(a_0, a_1)$ are: (a) (9.85714, 7.14286); (b) (7.08116,
2). The tongues are, from left to right: (a): 5:1,  33:8, 41:10, 4:1, 39:10, 35:9, 31:8,
27:7, 23:6, 19:5, and 34:9; (b): 5:1, 29:7, 33:8, 37:9, 41:10, 4:1, 39:10, 35:9, 31:8, and
27:7}
\end{figure}

\end{center}

\vspace{-3em} Fig.\ \ref{fig:tonguesdetailed} plots the results
obtained for two of the parameter sets. We only plotted those
tongues whose horizontal widths were larger than 0.00026 for at
least one value of the coupling $B$ (an arbitrary cutoff chosen
by trial and error, in order to plot neither too many nor too
few tongues) \cite{foot3}. We can see in that figure the vast
variety of synchronization regions that the system yields.
There are even some on this Fig.\ with $m=10$, the highest
value of $m$ that we trailed. As we reduce the lower cutoff of
the width criterion, many new tongues arise in the plot. The
effect of synchronization with the external action can be
inferred from the fact that these tongues are not vertical
lines, but curved.

Furthermore, comparison of Figs.\ \ref{fig:tonguesdetailed}(a) and
\ref{fig:tonguesdetailed}(b) shows that the tongues become closer
to vertical straight lines for smaller $a_1$, implying that
synchronization is more and more a matter of fine tuning of the
external driving frequency rather than of the interaction of the
rotator  with the external action. This phenomenon provides
further confirmation of our hypothesis that a strong dependence of
instantaneous frequency on angle (for the uncoupled rotator) helps
high-order synchronization.

\subsection{Competition among the tongues}\label{sec:competition}

The curving of the high-order tongues in Fig.\
\ref{fig:tonguesdetailed}(a) can be attributed to the
occurrence of \emph{competition among tongues}: in this case
the 4:1 tongue widens as the coupling increases and ``pushes''
the high-order tongues. Our circle map $h$ is always
invertible, because equation (\ref{rotatorforced}) can be
integrated with reversed time. Therefore, Lemma 1.5 in
\cite{boyland} says that the Arnold tongues cannot overlap.
That is the basis on which we argue that there must be
competition among the tongues.

\begin{center}
\begin{figure}[t!]
$$\begin{array}{c}
\includegraphics[height=0.48\textwidth, angle=270]{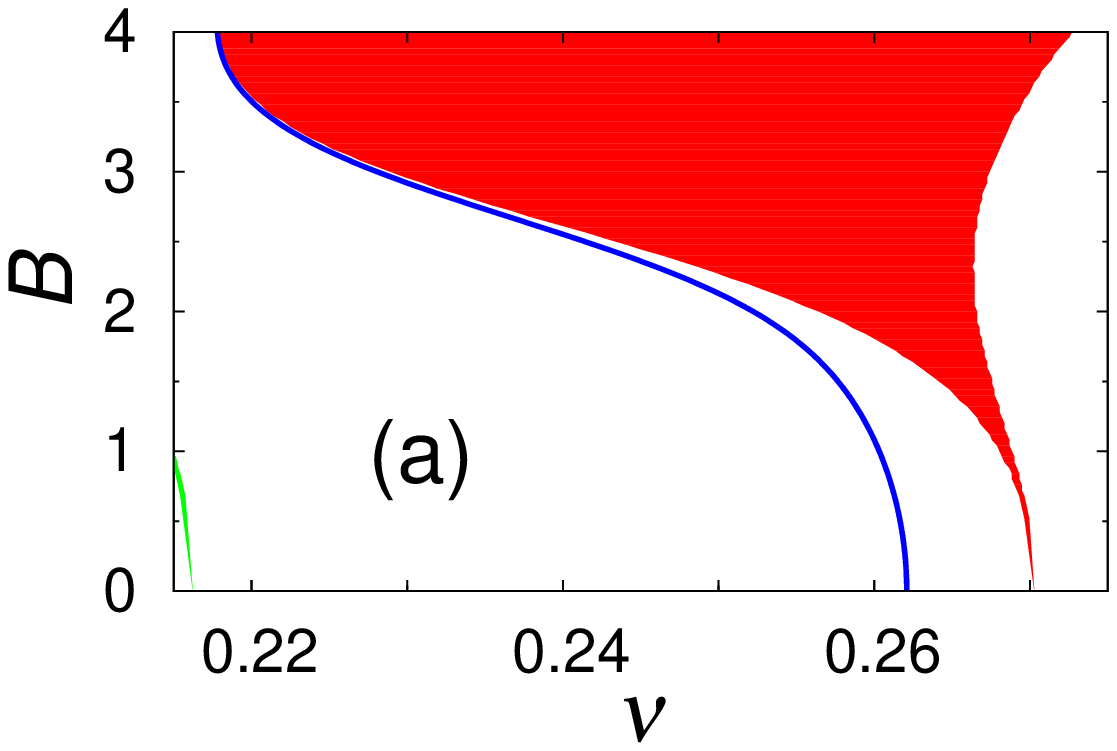}\\
\includegraphics[width=0.48\textwidth]{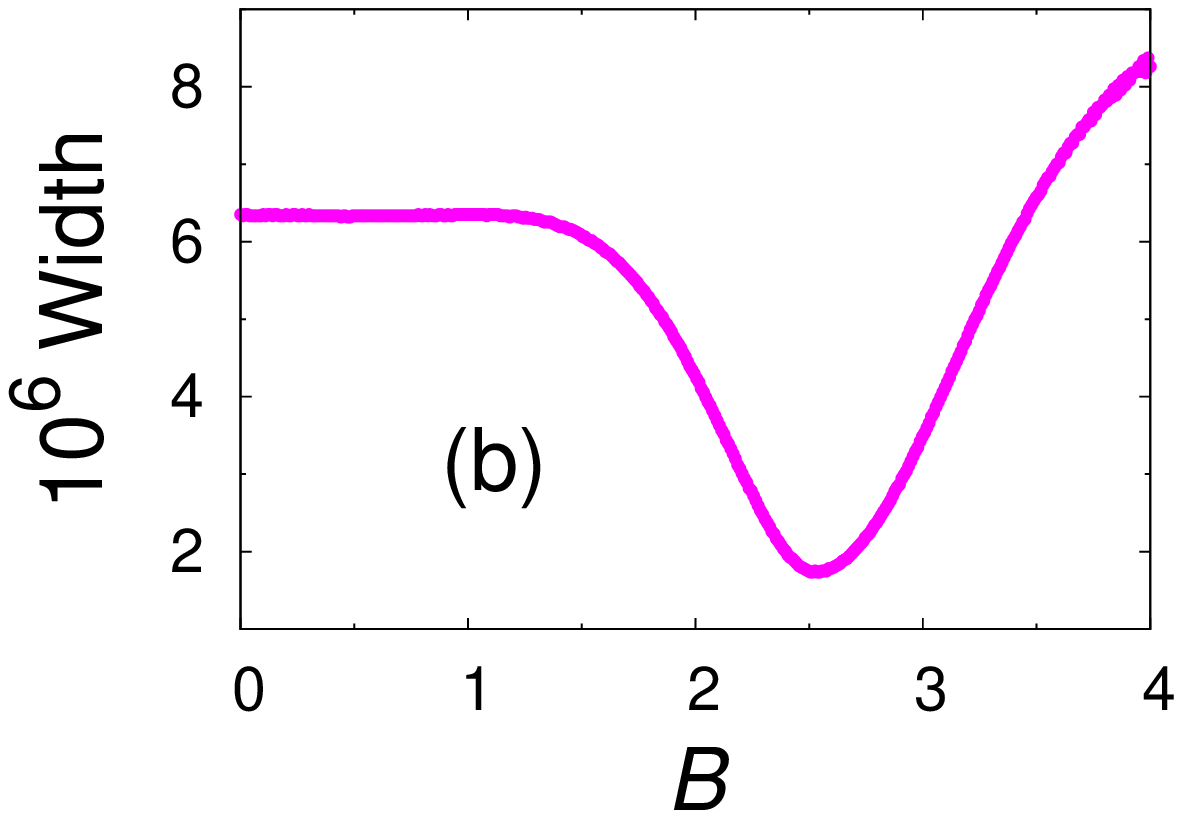}
\end{array}$$
\caption{\label{fig:2superdetailed} (colour online)
(a) Tongues 4:1  (right) and 33:8 (second from the right) for parameters
$(a_0, a_1)$=(9.85714, 7.14286). For clarity, the width of the 33:8 tongue in
(a) is shown as larger than it really is: this plot is just intended to show
the location. (b) Width of the 33:8 tongue, plotted as a function of the
driving amplitude $B$. The location of the 33:8 tongue in (a), and its
width in (b), were obtained by trailing the bifurcations in their
``flexible'' meaning, using a threshold of 0.001.
}
\end{figure}
\end{center}


We note here several signs of such competition. First, the 37:9
tongue appears in Fig.\ \ref{fig:tonguesdetailed}(b) but not in
\ref{fig:tonguesdetailed}(a), because it is narrower there. We
can interpret this effect as 33:8 and 41:10 \emph{squeezing} it
for the set of parameters used in \ref{fig:tonguesdetailed}(a).

In the same way, the  widths for 33:8 and 41:10 obtained via
the synchronization index in Fig.\ \ref{fig:tonguesdetailed}(a)
are of the order of $10^{-4}$ at $B\to 0$, and of the order of
$10^{-5}$ at $B=3$. Nevertheless, the width of the
synchronization region as $B\to 0$ must tend exactly to zero.
As indicated above, the computation of Arnold tongues via a
synchronization index (using (\ref{synchindex}) in our case) is
less accurate. Because of this evident contradiction between
the obtained width of the tongue as $B\to 0$ (width bigger than
zero) and the actual value of the width (strictly zero), we
need a more defined approach. The first try was to obtain the
bifurcation points with the first programme but it was
impossible: although we set a high precission, we failed to
find any zeros for the function $h_{nm}$.

That is why we are introducing the third method of obtaining
approximately Arnold tongues for tricky cases like this one:
\emph{trailing the bifurcations in their ``flexible'' meaning}.
Instead of considering that we are inside the tongue when
$h_{nm}$ has a stable zero, we relax the condition and just ask
$h_{nm}$ to have an ``almost (half stable) zero'', defined as a
point where $h_{nm}$ takes an absolute value smaller than a
threhold (set here to be 0.001). It means that, if we measure
the angle of the rotator at the times when the external action
has a maximum, this angle will stay for a long time around this minimum (or rather, minimum in absolute
value). What happens afterwards depends on the behaviour of $h_{nm}$. If $h_{nm}$ increases quickly (in absolute value) when we separate from the minimum, the angle will slip and come back promptly to the minimum (modulus $2\pi$). If $h_{nm}$ is still very small when we separate from the minimum, the angle will have a slow drift when iterating $h_{nm}$ so, in practice, the system will be regarded as synchronized, unless the measurements are very precise.

The tongue 33:8 obtained this way is plotted in Fig.\
\ref{fig:2superdetailed}(a), together with the 4:1 tongue
obtained before. In Fig.\ \ref{fig:2superdetailed}(b) we
plotted the width of the 33:8 tongue as a function of the
driving amplitude $B$. Because of the ``flexible'' way of
obtaining the tongue, its width tends to a value bigger than zero as $B\to 0$. The explanation is simple:
without any driving ($B=0$) there is no interaction, so the only possibility of having an apparent synchronization -- not true synchronization, as there is no interaction --  is to tune the external frequency to the value $m\nu_0/n$. This frequency is the only point of the Arnold tongue in the limit $B\to 0$. The function  $h_{nm}$ is there constantly equal to zero. If we make the external frequency slightly different from $m\nu_0/n$, with $B=0$, the values returned by the function $h_{nm}$ will still be very small, so many of them will be bellow the threshold; that is the reason why the obtained width of the tongue in the limit $B\to 0$ is bigger than zero.
Of course, this ``flexible'' method is not completely
accurate, and it gives the apparent contradiction above. Nevertheless, we are using this method only to obtain qualitative
conclusions, by \emph{comparing} results all obtained
with the same method. So we believe that the conclusions remain
valid.

As we can see from Fig.\ \ref{fig:2superdetailed}(b), the width
of the 33:8 tongue is more or less constant for $B$ near zero,
and then it starts dropping at around $B=1.5$. Looking at Fig.\
\ref{fig:2superdetailed}(a), we may interpret $B=1.5$ as the
point where the 4:1 tongue  has approached 33:8 enough to start
squeezing it. The width of 33:8 reaches a minimum at $B=2.5$
and then it increases. In Fig.\ \ref{fig:2superdetailed}(a),
$B=2.5$ is more or less the point where 4:1 switches from being
concave (concave from outside) to convex, so it is consistent
with the fact that the width of 33:8 starts increasing at this
point: let us say that, before this point, 4:1 was ``invading''
33:8; from this point, 33:8 ``starts recovering from 4:1's
invasion'', so 33:8 gains space (width) and 4:1 switches from
being concave to be convex (of course, there are,
mathematically speaking, an infinite number of tongues between
33:8 and 4:1, but 4:1 is the most dominant in the area so it is
the main influence).

The cases discussed above reveal clear signs of the effect of
competition among tongues; the existence of such competition
was predicted at the outset on strictly theoretical grounds.

\begin{center}
\begin{figure}[b!]
\includegraphics[width=0.48\textwidth]{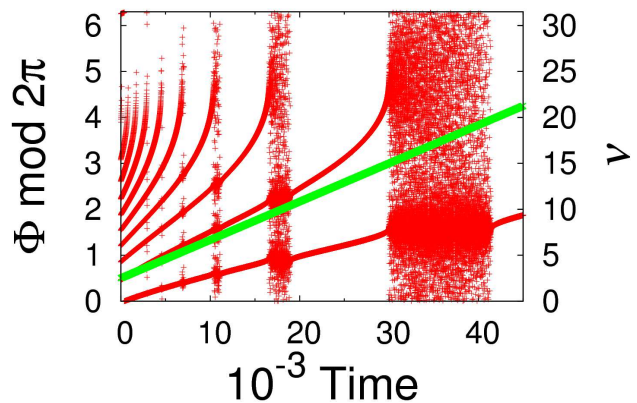}
\caption{\label{fig:changelin} (colour online). An $n$:1
synchrogram for the forced rotator while varying the driving
frequency. Several $n$:1 synchronization epochs can be observed.
The straight line plots the instantaneous frequency of the external
driving (right-hand ordinate axis).}
\end{figure}
\end{center}

\begin{center}
\begin{figure}
$$\begin{array}{c}
\hspace{-1em}\includegraphics[width=0.48\textwidth]{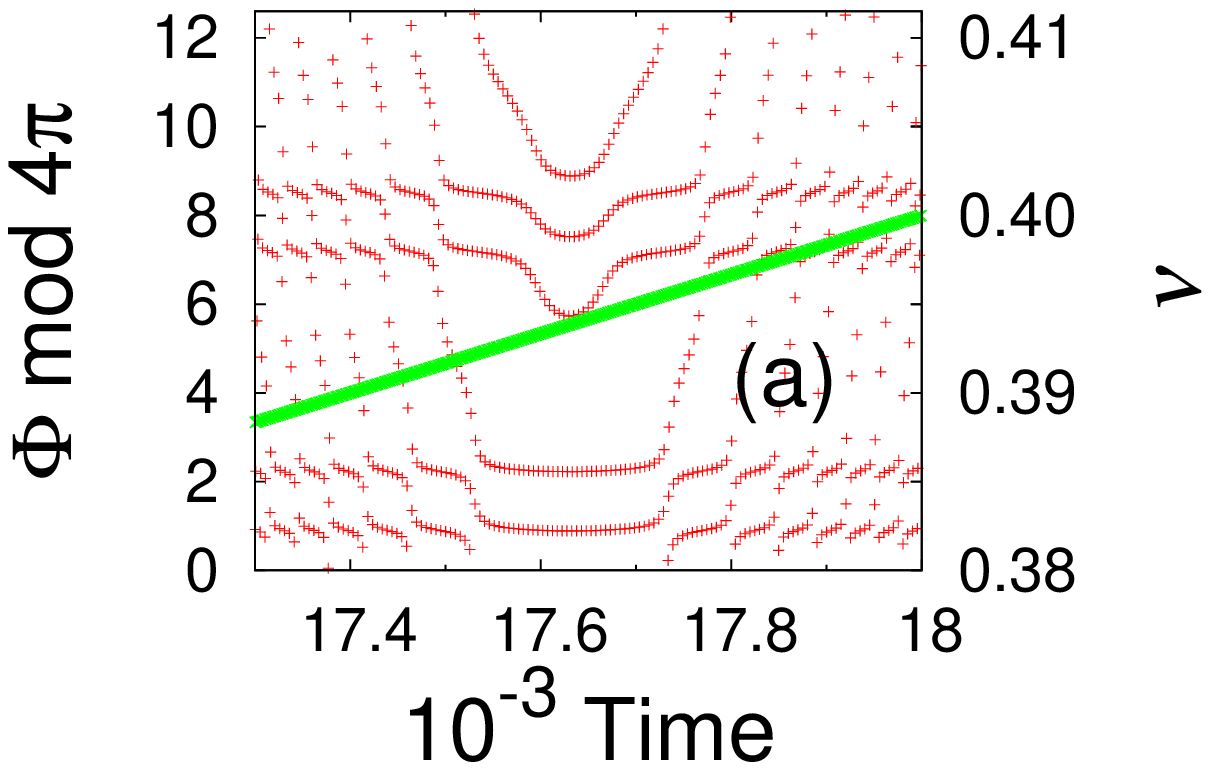}\\
\hspace{-1em}\includegraphics[width=0.48\textwidth]{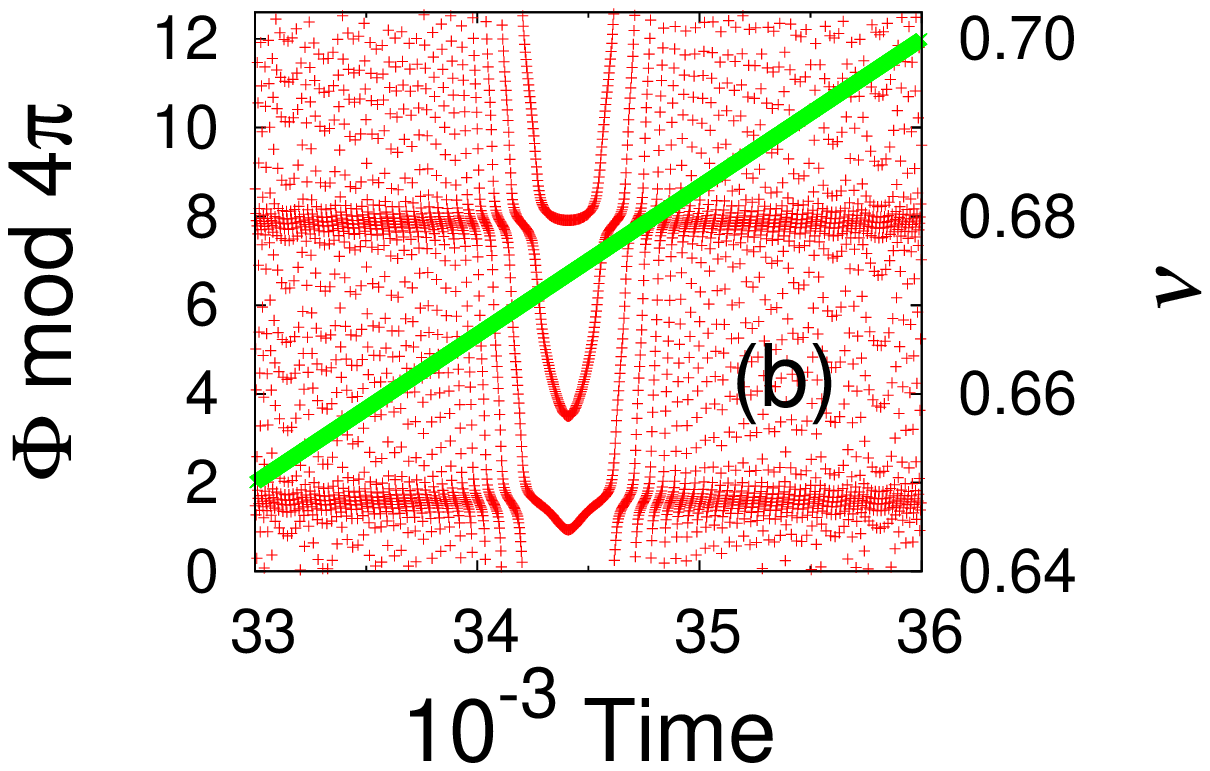}
\end{array}
$$
\caption{\label{fig:changelin2} (colour online). Sections of an
$n$:2 synchrogram for the forced rotator while varying the driving
frequency, illustrating a 5:2 synchronization epoch in (a) and a 3:2
epoch in (b). The straight line plots the instantaneous frequency of the
external driving (right-hand ordinate axis).}
\end{figure}
\end{center}

\vspace{-3em}

\section{Time variability and transitions between different synchronization epochs}
\label{sec:transitions}

Transitions in time between states of different $n$:$m$
synchronization ratios have often been observed experimentally
\cite{schafer, anetarats, bartsch, anetabracic}. We now consider in turn two
mechanisms that may give rise to such transitions in our system
(\ref{rotatorforced}): (i) variability of the driving frequency,
and (ii) low-frequency noise. The latter is mathematically
equivalent to slow variations of the autonomous frequency of the
rotator. Both mechanisms correspond to time variability, and they
can coexist.

\subsection{Variability of the driving frequency}

We again simulated the system (\ref{rotatorforced}), but
modified such that we replaced $B\sin(\omega t)$ with
$B\sin(\Phi(t))$, and varied the instantaneous frequency
$\dot\Phi(t)$ linearly in time
\begin{equation}\label{variarphi}
 \dot\Phi(t)=a\, t+b;\qquad \Phi(t)=\frac{1}{2} a\, t^2+b\, t.
\end{equation}

\noindent The choice of a linear
variation was arbitrary, in the interests of clarity.
With the parameters of (\ref{rotatorforced}) chosen
to be $(a_0, a_1)$=(9.85714, 7.14286), $B$=4, we slowly swept
$\dot\Phi(t)/2\pi$ from 0.1 to 1.3 in a total time of 72,000.
Fig.\ \ref{fig:changelin} shows $n$:$1$ synchronization epochs
corresponding to the times when the external instantaneous
frequency is inside an Arnold tongue. As $n$ decreases, the
synchronization epochs last longer, and the transition regions
widen. That is because, as mentioned above, the Arnold tongues
are then wider and further separated. The $n$:$2$ epochs are
also observed from the synchrograms: Fig.\ \ref{fig:changelin2}
shows $5$:$2$ and 3:2 synchronization epochs.

\begin{center}
\begin{figure}
$$\begin{array}{c}
\includegraphics[width=0.48\textwidth]{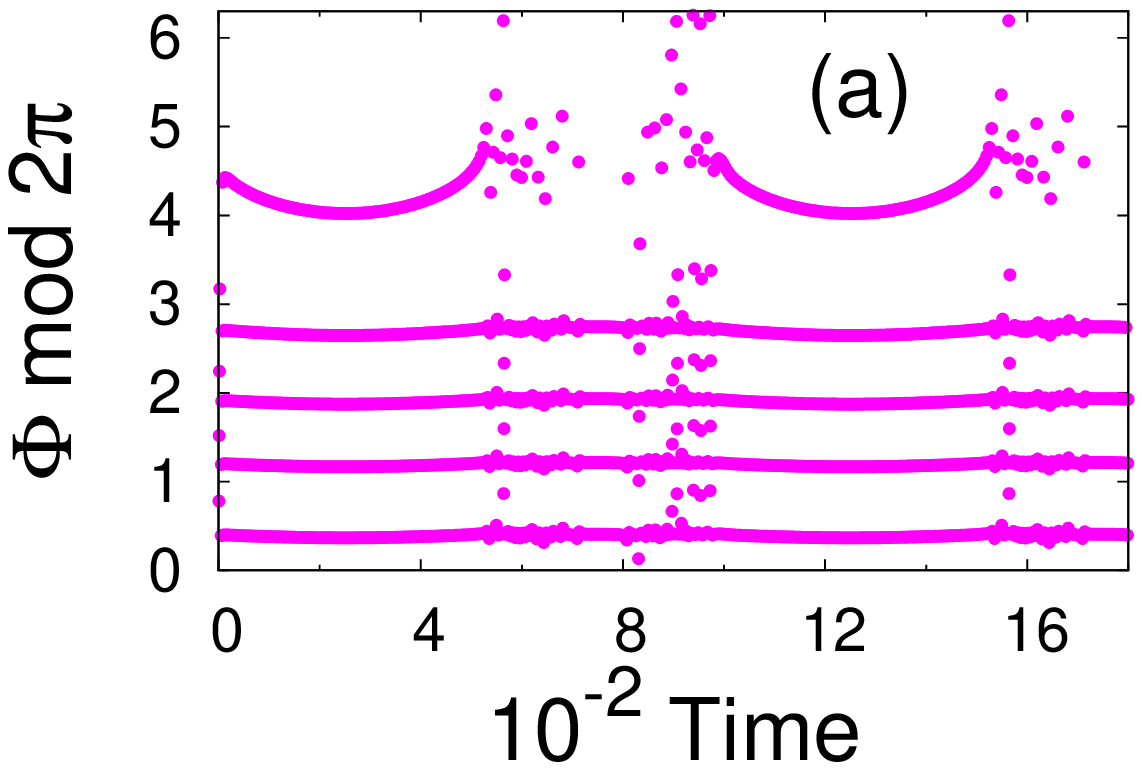}\\
\includegraphics[width=0.48\textwidth]{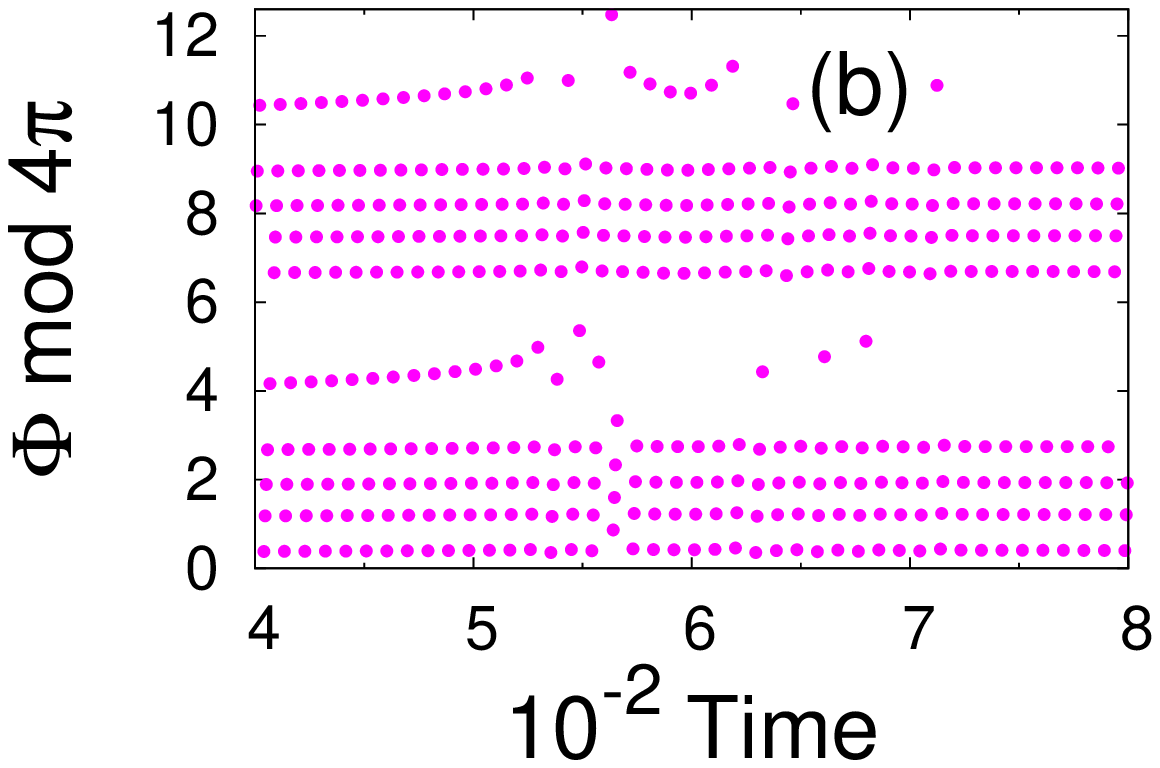}
\end{array}
$$
\caption{\label{fig:low1} (a) Transitions between 5:1 and 4:1
synchronization states, caused by the added low frequency
component. (b) A  9:2 epoch for time $\sim$600 in the
transition between 5:1 and 4:1.}
\end{figure}
\end{center}

\vspace{-3em}

For a Josephson junction, transitions of this kind could be
induced by variation of the frequency of the harmonic component of
the current driving the system. Where we regard
(\ref{rotatorforced}) as a simple model for the heart driven by
the respiration, the transition mechanism would correspond to
variation of the respiration rate. Although this rate can be
consciously controlled \cite{pacedresp}, people do not do so most
of the time, and so the small natural variations in respiration
rate can produce transitions in the sychronization ratio between
it and the heart.

\subsection{Low-frequency noise}

The second transition mechanism considered here arises from the
effect of the low-frequency noise that is often present in
physical, biological and economic systems. For instance, flicker
noise \cite{flicker} occurs in almost all electronic devices.
Another example is the cardiovascular system, where there are
known to be metabolic, neurogenic and myogenic \cite{anetageneral}
low-frequency oscillatory processes.

We therefore added to our system an extra low-frequency harmonic
component, simulating Eq.\ (\ref{rotatorforced}) with $(a_0,
a_1)$=(9.85714, 7.14286), $B$=4, $\omega/2\pi$=0.213, with a term
$L_1 \sin(\omega_1\, t)$ added to the RHS, $L_1$=0.1 and
$\omega_1/2\pi$=0.001. This value of $\omega_1$ may appear low
compared to some natural systems, but the results we will obtain
remain valid for larger $\omega_1$ provided it is small compared
with the frequencies of the other processes. Without adding the
low frequency component, we would be inside the 5:1 tongue and the
synchrogram would consist of 5 perfect horizontal lines. As shown
in Fig.\ \ref{fig:low1}(a), the corresponding simulation exhibits
transitions between 5:1 and 4:1. Note that we have added a
harmonic low-frequency component; with low-frequency noise, the
sequence of transitions will not be predictable in the manner seen
here. In Fig.\ \ref{fig:low1}(b) we see a 9:2 epoch inside the
transition region between the two longest epochs.

This second mechanism is mathematically equivalent to a slow
variation of the intrinsic frequency of the rotator. This can be
seen by considering the low-frequency component to be absorbed
inside $a_0$, yielding an $a_0$, and therefore a $\nu_0$, that
slowly vary with time. In nature, however, the picture is
completely different, as the low-frequency noise comes from
\emph{outside} the rotator.

\subsection{Time variability and synchronization}

Although we can still recognise synchronization epochs from the
synchrograms by sight, time variability -- the origin of the
two transition mechanisms discussed above -- means that
synchrograms no longer consist of perfect horizontal lines. The
test via a synchronization index, often used when tackling real
experimental data, must fail, even inside a small time window,
unless we take a less strict threshold for the index. Our
result is in agreement with that obtained in \cite{anetabracic}
by analysis of experimental data. In short, time variability
interferes with synchronization; at least for our system and
with the index (\ref{synchindex}) as a quantification of
synchronization.  A more detailed and quantitative study of the
influence of time variability is in progress and will be
reported elsewhere.

\section{Conclusions}\label{sec:conclusions}

We have shown that the driven rotator Eq.\
(\ref{rotatorforced}) yields a vast variety of Arnold tongues.
Our results are applicable to the several systems described by
this equation, including the overdamped pendulum and the
overdamped Josephson junction. Because the synchronization
regions can be relatively wide, and because (\ref{ourrotator})
is written in terms of continuous variables, the equation may
be useful in modelling the wide range of high-order
synchronization phenomena observed in nature, such as in the
cardiorespiratory interaction. Although we investigate a
specific system, and although the deep mechanisms for
synchronization are not revealed by our studies in this paper,
two of the results that we have obtained are very interesting
and likely to be useful in the quest for those deep mechanisms:
we showed that a strong dependence of the instantaneous
frequency on the angle helps high-order synchronization; and we
also identified and studied the phenomenon of competition among
tongues. Finally, we have discussed and explored two mechanisms
of transition between different synchronization states.

\begin{acknowledgments}

We thank A. Bahraminasab, A. Pikovsky and M. Rosenblum for useful discussions.

This work was supported by EU Project BRACCIA \cite{foot4}.

\end{acknowledgments}


\begin{thebibliography}{99}
\bibitem{foot1} Note that ``rotator'' and ``angle'' are
    different  from``phase oscillator'' and ``phase''. The equation of motion
    of a phase oscillator in the limit cycle is, by definition
    of phase, $\dot\phi=\omega$ \cite{kuramoto, arkadybook}.
\bibitem{foot2} In cardiovascular system recordings,
    the respiratory signal looks typically like that from a
    harmonic oscillator plus higher harmonics. On the other
    hand, it is known that the coupling from respiration to
    heart is stronger than vice versa \cite{directionality}.
\bibitem{foot3} To plot Fig.\ \ref{fig:tonguesdetailed} we
    removed a small artifact:
    inside the wide tongues 5:1 and 4:1, some regions of
    multiples with the same ratio appeared. For example, inside
    the 5:1 tongue, some areas with 15:3, 20:4, etc. were
    obtained. These regions were merged by hand into the wide
    tongue for two reasons: first, because visual inspection of
    the plot of $\theta(t)$ together with the plot of the
    external forcing revealed that it was actually 5:1 (or 4:1)
    synchronization; second and more important, because
    according to \cite{katok, boyland} the characteristic
    parameter of synchronization in a circle map is the
    rotation number $n/m$.
\end{thebibliography}
\end{document}